\documentclass[10pt,conference]{IEEEtran}
\IEEEoverridecommandlockouts
\usepackage{cite}
\usepackage{amsmath,amssymb,amsfonts}
\usepackage{algorithmic}
\usepackage{graphicx}
\usepackage{textcomp}
\usepackage{xcolor}
\def\BibTeX{{\rm B\kern-.05em{\sc i\kern-.025em b}\kern-.08em
    T\kern-.1667em\lower.7ex\hbox{E}\kern-.125emX}}

\usepackage{ftnright}
\usepackage{xspace}
\usepackage{amsmath,amsfonts}
\usepackage{graphicx}
\usepackage{textcomp}
\usepackage{xcolor}
\usepackage{soul}
\usepackage{xspace}
\usepackage{url}
\usepackage{enumitem}
\usepackage{colortbl} 
\usepackage{hyperref}
\usepackage{cleveref}
\usepackage{listings} 
\usepackage{graphicx}
\usepackage{tcolorbox}
\usepackage{hyperref}
\hypersetup{
    colorlinks=true,
    linkcolor=blue,
    filecolor=magenta,      
    urlcolor=cyan,
    pdftitle={Overleaf Example},
    pdfpagemode=FullScreen,
    }
\usepackage{subcaption}

\crefformat{section}{§#2#1#3}
\Crefformat{section}{§#2#1#3}
\newcommand*{\eg}{e.g.\@\xspace} 
\newcommand*{\ie}{i.e.\@\xspace} 

\newcommand{\code}[1]{\texttt{#1}}

\newcommand{\tool}{\textsc{SimShadow}\xspace}
\newcommand{\bench}{\textsc{MQTBench}\xspace}
\newcommand{\qasm}{\textsc{QASM}\xspace}
\newcommand{\qiskit}{\textit{Qiskit}\xspace}
\newcommand{\cirq}{\textit{Cirq}\xspace}
\newcommand{\ibm}{\textit{IBM Boston}\xspace}
\newcommand{\qhtwo}{\textit{Quantinuum H2}\xspace}

\makeatletter
\newcommand{\linebreakand}{%
  \end{@IEEEauthorhalign}
  \hfill\mbox{}\par
  \mbox{}\hfill\begin{@IEEEauthorhalign}
}
\makeatother

\begin{document}

\title{Toward Live Noise Fingerprinting in Quantum Software Engineering}

\author{
    \IEEEauthorblockN{Avner Bensoussan}
    \IEEEauthorblockA{\textit{King's College London}, UK \\}
    \and
    \IEEEauthorblockN{Elena Chachkarova}
    \IEEEauthorblockA{\textit{King's College London}, UK \\}
    \and
    \IEEEauthorblockN{Karine Even-Mendoza}
    \IEEEauthorblockA{\textit{King's College London}, UK \\}
    
 \linebreakand
    
    \IEEEauthorblockN{Sophie Fortz}
    \IEEEauthorblockA{\textit{Inria, Univ. Rennes, CNRS, IRISA}, France \\}
    \and
    \IEEEauthorblockN{Vasileios Klimis}
    \IEEEauthorblockA{\textit{Queen Mary University of London}, UK \\}
    \and
    \IEEEauthorblockN{Mohammad Reza Mousavi}
    \IEEEauthorblockA{\textit{King's College London}, UK \\}
}

\maketitle

\begin{abstract}


Contemporary quantum computers are inherently noisy, posing significant challenges for the development and testing of quantum software. Simplified or outdated noise assumptions can lead to incorrect assessments of program correctness, obscure debugging, and hinder cross-platform portability, creating a critical quantum software development gap. 
Providing accurate, practical noise characterisation is challenging as traditional reconstruction methods scale exponentially and rapidly become outdated.

In this vision paper, we address this gap via a novel \textit{classical shadow tomography}-based pipeline, \tool{}, enabling efficient, continuously updatable \textit{noise fingerprinting} from empirical observations, suitable for integration into quantum software development workflows, including testing and validation.
We prototyped the pipeline to investigate fingerprints' ability to capture structured, interpretable noise and cross-platform discrepancies affecting quantum programs' behaviour to support realistic testing and debugging in future tools.
Our evaluation with \qiskit and \cirq under widely used hardware-informed profiles, \ibm and \qhtwo, shows fingerprints exhibit channel-specific structure and yield interpretable heatmaps. We observed systematic cross-platform discrepancies under matched noise configurations, quantified by large Frobenius distances at a fraction of full tomography cost. On 69 \bench programs, larger fingerprint differences correlate with output distributions divergences, highlighting threats for testing and cross-platform debugging tasks.
\end{abstract}


\section{Introduction}
\label{sec:into}

In contemporary Quantum Software Engineering (QSE) \cite{QSERoadmap2025,DESTEFANO2024107525}, activities such as development, compilation and debugging face a unique obstacle: sensitivity to the ``noise'' of quantum hardware.  Quantum programs are affected by fragile qubits whose behaviour is disrupted by environmental fluctuations and imperfect control \cite{nielsen2010quantum,10123493,bharti_noisy_2022,piskor2025simulationbenchmarkingrealquantum}.  In quantum ecosystems (\eg{} \qiskit or \cirq), simulators are indispensable for QSE, providing controlled environments for a variety of tasks and enabling low-cost experimentation \cite{ibm_pricing,azure_quantum_pricing}.  Simulators such as those for \qiskit \cite{qiskit2024} and \cirq \cite{Cirq_Developers_2025} must capture not only the ideal logic of algorithms but also the noisy, platform-specific imperfections of real hardware \cite{nielsen2010quantum,qiskit2024_aer_exact_noisy_simulation,cirq_representing_noise_2025}. These imperfections are abstracted into \emph{noise models}, which are often inaccurate and out-of-sync with respect to real hardware \cite{Maschek_2025, BravoMontes_2024}. Recent work has focused on improving these models \cite{Maschek_2025}, highlighting their central importance for reliable and effective QSE \cite{10821340,10.1145/3297858.3304075,10.1145/3428198,10682972}.

Yet, a critical \emph{conceptual gap} exists: there is a significant lack of accurate and practically actionable noise models.  Although platforms offer typical noise channels (\eg{} ``depolarising noise''), the underlying semantics of their implementations differ significantly. This undermines the role of simulators in QSE for evaluating quantum software behaviour in noisy environments and represents a fundamental, unaddressed challenge for building reliable and transferable quantum software.

We propose \emph{efficient noise fingerprinting} as a new vision for addressing this challenge. 
We adapt \emph{classical shadow tomography}~\cite{huang2020predicting}, a lightweight quantum mechanics technique, as a software engineering analysis tool to generate rich, descriptive signatures of a simulator's noise profile, thereby avoiding the exponential cost of accurately reconstructing the quantum state under noise.  
We investigate this paradigm empirically. As a proof of concept, we implement the pipeline, \tool, and study two research questions:
\begin{enumerate}[nosep,noitemsep,leftmargin=25pt,itemsep=0pt, topsep=0pt]
    \item[\textbf{RQ1}] 
    Do fingerprints reveal and differentiate noise behaviours within a platform?
    \item[\textbf{RQ2}] 
    Can \tool{} detect and quantify differences between platforms with identical noise configurations?
\end{enumerate}
Investigating this enables downstream applications such as fault localisation, differential testing (distinguishing implementation bugs from noise-induced variation), and automated testing under realistic noise conditions.
Preliminary Qiskit-Cirq comparison yields interpretable heatmaps, revealing systematic implementation differences, with fingerprints exhibiting channel-specific structure.
Across six configurations (3 noise channels × 2 hardware-informed profiles, \ibm and \qhtwo), the platform-level fingerprints Frobenius distance ($\approx 0.0044$--$0.0056$) is an order of magnitude greater than the bootstrap uncertainty ($\approx 0.0004$), while remaining substantially more efficient than full process tomography.


Our work is the first to close the aforementioned conceptual gap by leveraging classical shadow tomography to enable efficient noise fingerprinting for quantum software platforms in the QSE context. We envisage that our vision for practical noise fingerprinting will be transformative for other areas, including noise-aware compilation, transpilation, cross-platform validation, verification, and error mitigation.

\vspace{0.2 cm}
\noindent\textbf{Data Availability. } Code, fingerprints images, results, and testing notebooks (\cref{sec:threats}) are openly available at \cite{tool:zenodo}.

\begin{figure}[t!]
\centering
\includegraphics[width=1.01\columnwidth]{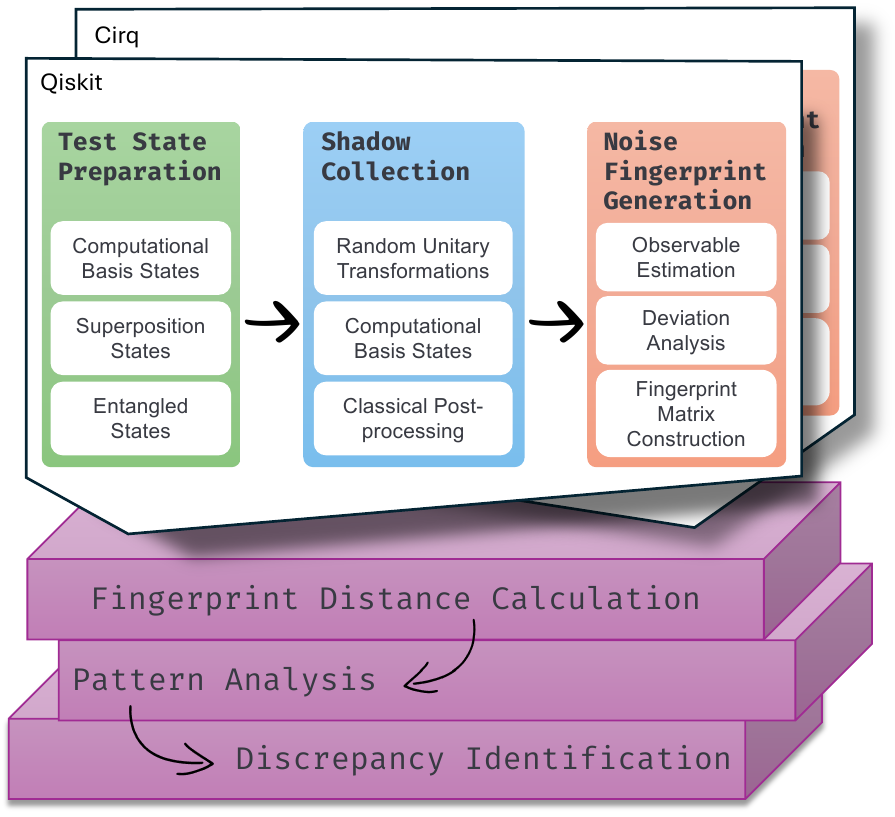}
\caption{\tool Architecture}
\label{fig:architecture}
\end{figure}

\section{Background \& Related Work}
\label{sec:back}
\noindent\textbf{Basics of Quantum Measurement. } \cite{nielsen2010quantum}. %
A (pure) quantum state is represented by a ket $\vert \psi \rangle$ in a complex Hilbert space; its dual bra is $\langle \psi \vert = (\vert \psi \rangle)^\dagger$; for a single qubit with computational basis $\{\vert 0 \rangle, \vert 1 \rangle\}$, the general state is a \textit{superposition} $\vert \psi \rangle = \alpha \vert 0 \rangle + \beta \vert 1 \rangle$, where \( \alpha \) and \( \beta \) are complex amplitude coefficients (\eg $|+_y\rangle = (|0\rangle + i|1\rangle)/\sqrt{2}$).  An \emph{observable} $O$ is a Hermitian matrix whose eigenvalues are the possible measurement outcomes; the expected result of measuring $O$ on $\vert \psi \rangle$ is $\langle \psi \vert O \vert \psi \rangle$.  
Density matrix $\rho = |\psi\rangle\langle\psi| \in \mathbb{C}^{2^n \times 2^n}$ represents an $n$-qubit system general state and scales exponentially in $n$.

\vspace{0.1 cm}
\noindent\textbf{Error Mitigation. }%
Quantum computations are affected by noise from imperfect control, decoherence, and measurement errors \cite{nielsen2010quantum}. Noise is studied on \textit{quantum hardware}, where device-specific effects can be time-dependent, correlated, and non-Markovian or in \textit{simulation}, where errors are modelled via \textit{abstract quantum channels} (\eg depolarising or amplitude-damping) but fail to reproduce accurately the statistical behaviour observed on devices.  \textit{Error mitigation} seeks to reduce noise impact without full quantum \textit{error-correction} overhead, 
\eg \cite{temme2017error,endo2018practical}.  
Shadow tomography and classical shadows, used alongside error mitigation methods, utilise randomised measurements to estimate many observables without full state reconstruction.
\textit{Shadow tomography} \cite{10.1145/3188745.3188802,10.1145/3313276.3316378} retains a list of measurement vectors for implicit density matrix approximation.  \textit{Classical shadows} \cite{huang2020predicting} compacts this list into \emph{classical} data structures for prediction of observables.  
\tool{} uses shadow-inspired estimation but differs from both: it stores no shadow representation and maps outcomes directly to deviation signatures tailored specifically to QSE.

\vspace{0.1 cm}
\noindent\textbf{Quantum Software Testing }is challenging, with techniques such as property-based~\cite{honarvar2020qsharpcheck,pontolillo2025qucheck}, mutation~\cite{wang2022mutationbased} and metamorphic~\cite{abreu_metamorphic_2022, paltenghi_morphq_2023}. 
Benchmarks such as 
Bugs4Q~\cite{zhao2023bugs4q} 
and MQTBench~\cite{quetschlich2023mqtbench} support testing and debugging studies.  Most approaches assume ideal or simplified noise models \cite{paltenghi_morphq_2023,fuzzq}, compromising test reliability. Noise-aware testing~\cite{pontolillo2025noiseaware} demonstrates the need for practical, 
noise characterisation to support robust testing, debugging, and cross-platform analysis.

\vspace{0.1 cm}
\noindent\textbf{Quantum Fingerprinting} was introduced by Buhrman et al.~\cite{buhrman_fingerprints} to encode inputs into exponentially small quantum states for equality testing.
We intend to use their framework in the future to establish formal properties of our approach.

\section{Live Empirical Noise Fingerprinting}
\label{sec:method}
We study the problem of undocumented discrepancies in quantum noise models by introducing \tool{} to create lightweight noise fingerprints to characterise noisy behaviour of a quantum simulator (\eg{} \qiskit's \texttt{\small AerSimulator}) as a live, dynamic signature for QSE tasks using two core ideas: 1) noise models fingerprinting and 2) quantum measurements.

\vspace{0.1 cm}
\noindent\textbf{Defining the Fingerprint. } (\cref{sec:paradigm} with formalism in \cref{sec:fingerprint:func}) %
To analyse these discrepancies, we introduce the notion of a noise fingerprint as a compact, rapidly updatable matrix of a simulator’s observed outcome deviations from ideal outcomes, distinguishing it from physics-oriented noise models for device-level diagnostics.  By probing reference states with a fixed set of Pauli observables and aggregating deviations, \tool obtains a descriptive matrix, making simulator differences explicit, measurable, and comparable across noises, simulators and platforms.

\vspace{0.1 cm}
\noindent\textbf{Analysis \& Application. }%
Fingerprints are then used in the analysis loop (\cref{sec:design}).  We applied it to \qiskit and \cirq, revealing systematic discrepancies in their simulator behaviour (\cref{sec:eval}); we outline broader applications in \cref{sec:agenda}. 

\subsection{The Noise Fingerprinting Paradigm}
\label{sec:paradigm}
Existing vendor-supplied noise models provide the closest available characterisation of device behaviour. Still, they are necessarily simplified, offer limited transparency and detail, and are updated only periodically (\eg{} every 24 hours). We advocate for a new paradigm centred on live, empirical and descriptive noise fingerprinting, designed to be transparent and accessible, with the following 3 key properties.

\vspace{0.1 cm}
\noindent\textbf{Empirical. }%
A fingerprint is generated from \emph{empirical data} by actively probing the target system (simulator or hardware) through measurement, not from a theoretical model or vendor's datasheet.  Here, fingerprints represent the \emph{actual}, ``de facto'' behaviour of the system, not the ``de jure'' behaviour described in its documentation.  Our approach grounds noise understanding in the observable reality, ``de facto'' system behaviour, unlike (\eg) process tomography~\cite{chuang1997prescription} and gate set tomography~\cite{Blume-Kohout2017}, which provide physically complete but exponentially costly noise characterisations impractical beyond a few qubits.

\vspace{0.1 cm}
\noindent\textbf{Live and Updatable. }%
Our fingerprint is \emph{live}: it can be generated quickly and updated repeatedly.  This is vital for hardware, where calibration drifts within hours, and for software, where fingerprints can be generated on demand for specific versions or commits in a CI/CD pipeline.  Prior QSE methodologies mainly focused on testing and benchmarking~\cite{DESTEFANO2024107525}, reporting on execution speed~\cite{miranskyy2020testing,property-based,fuzzq}.  Our approach captures fidelity and consistency by transforming noise from a static property into a dynamic, observable state of the system.

\vspace{0.1 cm}
\noindent\textbf{Descriptive and Actionable. }%
Randomised benchmarking~\cite{magesan2011scalable} yields only an aggregate fidelity score, revealing nothing about \emph{how} or \emph{why} simulators differ. We instead propose fingerprints as richer signatures: a \emph{descriptive, structured signature} as a matrix of deviations.  Its patterns may reveal the \emph{type}, \emph{strength} and \emph{asymmetries} of noise, making fingerprints \emph{actionable}. 
\looseness=-1

\subsection{Fingerprint as a Function}
\label{sec:fingerprint:func}
A \emph{configuration} $c$ specifies the ecosystem, backend (or hardware-informed profile), noise channel family and parameters, and software version/seed policy.  Given a fixed probe suite of $k$ reference states $\{|\psi_i\rangle\}$ and $m$ Pauli observables $\{O_j\}$, \tool defines the fingerprint $F(c)\in\mathbb{R}^{k\times m}$:
\begin{equation}
    F_{i,j}(c)=\widehat{\mu}_{i,j}(c)-\mu^{\mathrm{ideal}}_{i,j},
\qquad
\mu^{\mathrm{ideal}}_{i,j}=\langle\psi_i|O_j|\psi_i\rangle .
\end{equation}
Here $\widehat{\mu}_{i,j}(c)$ is the empirical estimator of $\mathbb{E}[O_j]=\mathrm{Tr}(\rho_c^{(i)} O_j)$, obtained by executing the noisy circuit for $|\psi_i\rangle$, rotating into the measurement basis of $O_j$, sampling $N$ shots, mapping outcomes to $\pm1$, and averaging.  This approach is \emph{shadow-inspired}, as it efficiently estimates selected Pauli expectations without reconstructing the full density matrix or storing a classical shadow representation (\S\ref{sec:back}).
\looseness=-1

\subsection{Shadow-Inspired Estimation}
\label{sec:design}
\tool targets deviation estimates rather than full physical reconstruction. Its fingerprints are intentionally less precise than physics-oriented tomography but are lightweight, interpretable, and well-suited for software analysis and testing.

Our \textit{protocol for generating and analysing a fingerprint} instantiates this paradigm through the following stages, illustrated in \autoref{fig:architecture}.

\vspace{0.1 cm}
\noindent\textbf{1. Prepare Reference States. }%
We prepare a small but diverse, well-characterised quantum states set $\{|\psi_i\rangle\}$, chosen so that different error mechanisms manifest distinctly in the fingerprint:
\begin{itemize}[nosep,noitemsep,leftmargin=10pt]
    \item \emph{Computational basis states} (\eg $|00\rangle, |11\rangle$), primarily sensitive to population errors such as amplitude damping.
    \item \emph{Superposition states} (\eg $|+\rangle|1\rangle, |0\rangle|+\rangle$), sensitive to coherence loss such as phase damping.
    \item \emph{Entangled states} (\eg{} a two-qubit GHZ state), necessary to probe correlated multi-qubit errors such as crosstalk.
\end{itemize}

\vspace{0.1 cm}
\noindent\textbf{2. Estimate a Selected Set of Observables. }%
After evolving each reference state under a configuration, $c$, we estimate expectation values for a selected set of Pauli observables $\{O_j\}$, including two-qubit correlators. Rather than reconstructing a density matrix, we directly estimate observables required for the fingerprint, capturing correlated noise effects efficiently.

\vspace{0.1 cm}
\noindent\textbf{3. Construct the Deviation Fingerprint. }%
For each reference state $|\psi_i\rangle$ (rows) and Pauli observable $O_j$ (columns), we assemble the fingerprint matrix $F(c)\in\mathbb{R}^{k\times m}$, where entry $F_{i,j}(c)$ corresponds to measuring $O_j$ on $|\psi_i\rangle$. Thus, $F(c)$ forms a state--observable grid of deviations from ideal behaviour, not a density matrix.
Ideal expectations are computed analytically; empirical estimates $\widehat{\mu}_{i,j}(c)$ are obtained without density matrix reconstruction.
We compare configurations using the Frobenius distance $\lVert F(c_A)-F(c_B)\rVert_F$, yielding a compact, quantitative measure of cross-platform divergence.

\vspace{0.1 cm}
\noindent\textbf{4. Algorithmic Analysis. }%
We analyse the resulting fingerprint matrices empirically using summary statistics and visual comparisons, as detailed in the next section.

\section{Emerging Results}
\label{sec:eval}
We generated 12 fingerprints, one per configuration, and compared them visually and through statistical distances and divergences to answer RQ1 and RQ2 (\cref{sec:into}).
\looseness=-1

\subsection{Experimental Setup}
\label{sec:eval:method}
\noindent\textbf{Configurations. }
The evaluation used 12 configurations (3 noise channels $\times$ 2 simulators $\times$ 2 profiles).
We evaluated \tool on \textbf{2} simulator ecosystems: \qiskit and \cirq, with \textbf{2} hardware-informed profiles: \emph{\ibm} (1Q gate-error proxy with damping mapped per $\sim$1Q gate step; see IBM’s Aer noisy simulation guidance~\cite{ibm_quantum_simulate_with_aer,ibm_quantum_build_noise_models}) and \emph{\qhtwo} (datasheet-anchored 1Q gate infidelity and depth-1 memory error; Quantinuum reports memory error rather than $T_1/T_2$~\cite{quantinuum_h2_product_datasheet,quantinuum_faq_doc,quantinuum_performance_validation_doc}).
We used \textbf{3} noise channels: \textit{depolarising}, \textit{amplitude damping}, and \textit{phase damping}, modelling respectively generic random errors, energy relaxation (a $T_1$-like decay $\lvert 1\rangle \rightarrow \lvert 0\rangle$), and pure dephasing (a $T_2$-like loss of coherence)~\cite{qiskit_aer_depolarizing_error_doc,qiskit_aer_amplitude_damping_error_doc,qiskit_aer_phase_damping_error_doc}. 

\vspace{0.1 cm}
\noindent\textbf{Probe suite. }  
We designed a protocol adapted from standard practices in quantum system characterisation~\cite{Cross2019, magesan2011scalable}.  We used $k=13$ reference quantum states prepared with standard minimal circuits, strategically chosen to be sensitive to a wide range of error mechanisms (\S\ref{sec:design}): \emph{$4$ computational basis states}, \emph{$8$ superposition states} and \emph{$1$ entangled state}. We measured $m=9$ two-qubit Pauli observables, covering all correlated errors per state, utilising a total of 117 state–observable pairs (13 states $\times$ 9 observables, \ie{} size of each fingerprint).

\vspace{0.1 cm}
\noindent\textbf{Statistical Robustness. }%
We ran $n{=}1000$ independent repetitions of the probe suite per configuration. We reported the mean fingerprint across repetitions with \code{\small shots=10{,}000} per state--observable measurement to reduce finite-shot sampling error, and as shot count being simulators divergence's prime factor \cite{fuzzq}. Furthermore, we bootstrapped $n$ runs: 2,000 resamples with replacement, and per resample computed mean \qiskit and \cirq fingerprints and their Frobenius distance, with std. dev. being std. err. of our point estimate (distance between original mean matrices).
\looseness=-1

\subsection{Threats to Validity}
\label{sec:threats}
\vspace{-0.1 cm}
\noindent\textbf{Internal validity:} 
To reduce the risk of discrepancies stemming from implementation or experimental environment, we developed the pipeline incrementally and validated intermediate steps via dedicated test notebooks before constructing the full flow. To minimise environment-induced variation, we pinned dependencies and ran all experiments in the same virtual environment (CloudLab c6525-25g \cite{cloudlab2019,cloudlab:hardware}).  

\vspace{0.1 cm}
\noindent\textbf{Conclusion validity:} 
Also, we utilised statistically robust setup (\cref{sec:eval:method}):
Expectation values lie in $[-1,1]$, hence fingerprint differences are in $[-2,2]$; the observed run-to-run variability is small relative to this range (std. dev.\ $\le 0.01076$). We used two fixed-parameter synthetic noise profiles, \emph{Low}\footnote{\ie $p_{\mathrm{dep}}=5\times10^{-4},\,\gamma_{\mathrm{AD}}=1\times10^{-4},\,\gamma_{\mathrm{PD}}=2\times10^{-4}$.} and \emph{High}\footnote{\ie $p_{\mathrm{dep}}=5\times10^{-3},\,\gamma_{\mathrm{AD}}=2\times10^{-3},\,\gamma_{\mathrm{PD}}=5\times10^{-3}$.}, to verify that the \emph{High} profile induces stronger decoherence and correspondingly larger cross-platform (\qiskit vs \cirq) discrepancies than the \emph{Low} profile; this was confirmed.
\looseness=-1

\subsection{Results} 
\label{sec:res}
\begin{figure}[t!]
\centering
\includegraphics[width=1.25\columnwidth,trim={1.4cm 7.4cm 5.5cm 0.6cm}, clip]{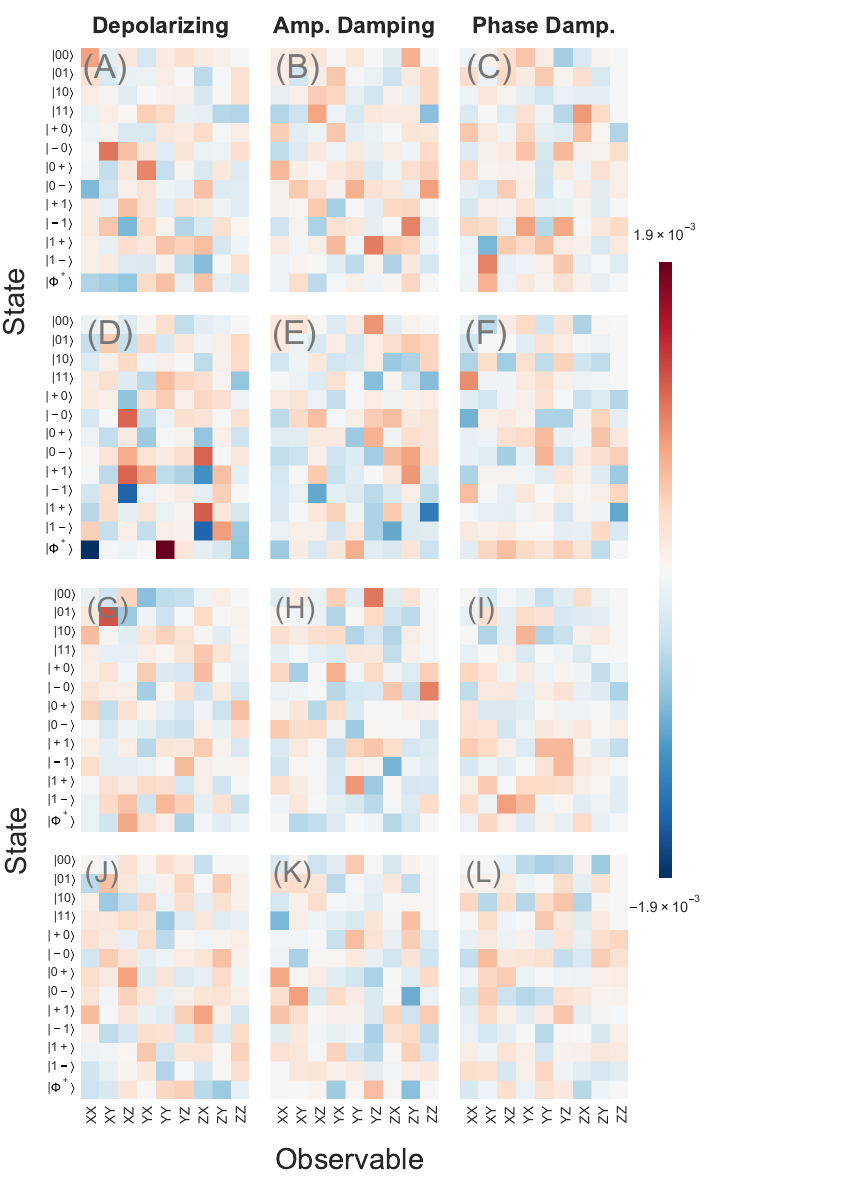} 
\caption{\textbf{\ibm vs \qhtwo.}} 
\label{fig:fingerprints}
\end{figure}

\Cref{fig:fingerprints} visualises mean fingerprints ($13\times 9$ matrix) with reference states as rows and two-qubit Pauli observables as columns (\eg{} $XX, XY, \ldots$). Entries report deviations from the corresponding ideal expectations. Rows group configurations by hardware-informed profile and platform: \ibm \qiskit (A--C) and \cirq (D--F); \qhtwo \qiskit (G--I) and \cirq (J--L).
Columns correspond to depolarising, amplitude-damping, and phase-damping noise channels.

\vspace{0.1 cm}
\noindent\textbf{RQ1: } Each noise channel exhibits a distinct visual structure, supporting noise-channel identification in \Cref{fig:fingerprints}. Within a fixed (profile, channel) pair, \qiskit and \cirq show clear, systematic differences. 
Nonetheless, we focus mostly on \ibm in this analysis, since its larger error magnitudes make the resulting signatures and trends easier to notice and interpret than on Quantinuum. 
We leave for future work the investigation of an optimal reference state per ecosystem.

\textbf{In (D)}, the depolarising noise results, the Bell state $|\Phi^{+}\rangle$ shows the largest deviations in $XX$, $YY$, and $ZZ$, expectedly as these correlators are ideally $\pm 1$ for $|\Phi^{+}\rangle$ and depolarisation uniformly shrinks all non-identity Pauli expectations towards zero. 
In addition, the product superposition states 
$|+0\rangle,|-0\rangle,|+1\rangle,|-1\rangle$ deviate strongly under $XZ$ and $|0+\rangle,|0-\rangle,|1+\rangle,|1-\rangle$ under $ZX$, because these are exactly the pairs for which $\langle XZ\rangle$ or $\langle ZX\rangle$ is ideally $\pm 1$.

\textbf{In (B) and (E)}, the amplitude damping experiments, largest $ZZ$ deviations occur for $|01\rangle$ and $|10\rangle$, consistent with relaxation driving single excitation states toward $|00\rangle$ and shifting $\langle ZZ\rangle$ from $-1$ toward $+1$; we observe pronounced changes for $|11\rangle$, as decay on either qubit first transfers population to $|01\rangle$ and $|10\rangle$ (which yield $ZZ=-1$) before relaxing to $|00\rangle$, producing a sizable net shift.
For the remaining amplitude damping results, the largest deviations occur in $ZX$, $XZ$, $ZY$, $YZ$, and $ZZ$ (3rd and last four col.), consistent with relaxation biasing $Z$ populations (captured by $ZZ$) and suppresses coherences when a qubit is probed in the $X/Y$ basis (captured by $XZ$, $ZX$, $YZ$, $ZY$); observables without a $Z$ component are less effected.

\textbf{In (H) and (K)}, \qiskit and \cirq show consistent behaviour for the $ZZ$ observable on $|01\rangle$, $|10\rangle$, and $|11\rangle$, with deviations only for states more sensitive to amplitude damping.

\textbf{In (C)}, for phase damping, we expect the largest effects on coherence dominated inputs, in particular the Bell state $|\Phi^{+}\rangle$, and the product superposition states $|+0\rangle,|-0\rangle,|+1\rangle,|-1\rangle$ under $XZ$ and $|0+\rangle,|0-\rangle,|1+\rangle,|1-\rangle$ under $ZX$, since dephasing suppresses off-diagonal coherences while leaving $Z$-basis populations comparatively unchanged; the deviations in panel \textbf{(C)} align with this expectation, with the lower half of the heatmap appearing more strongly coloured.

\begin{tcolorbox}[colback=gray!20, colframe=black,boxrule=1.0pt, before skip=5pt, after skip=5pt, left=1.9pt,
  right=1.9pt,
  top=1.9pt,
  bottom=1.9pt]
\textbf{RQ1 Answer:} 
Yes. Fingerprints show channel-specific, theory-consistent structure across states and observables. 
\end{tcolorbox}

\vspace{0.1 cm}
\noindent\textbf{RQ2: } 
Frobenius~\cite{Golub_VanLoan_1996} distances between \qiskit and \cirq mean fingerprints were (approx.)
for \ibm profile 0.0056 (depolarising), 0.0051 (amplitude damping), and 0.0049 (phase damping) and 
\qhtwo, 0.0046, 0.0051, and 0.0044, respectively.
Bootstrap resampling yields std. err. of approx. $0.0004$ for these distances, indicating that the observed differences are well above the uncertainty in estimating the mean distance (\eg{} $0.0056 \gg 0.0004$), confirming simulators exhibit distinct behaviours despite identical noise configurations.

For two configuration pairs (F–I and A–D, with the former exhibiting smaller fingerprint deviations than the latter), we evaluated portability on 69 \bench  \qasm programs\cite{quetschlich2023mqtbench} of 5-11 qubits (see list in \cite{tool:zenodo}).  All 4 configurations were executed with \code{\small shots=10{,}000}; \textit{Jensen–Shannon divergence} (JSD) was computed over 30 repeats per circuit (mean \& std.).
A–D exhibits slightly higher cross-configuration divergence on benchmark circuits (JSD 0.0728, std. dev. 0.1559) than I-F (JSD 0.0693, std. dev. 0.1364), supporting the link between fingerprint discrepancy and portability impact.

\begin{tcolorbox}[colback=gray!20, colframe=black,boxrule=1.0pt, before skip=5pt, after skip=5pt, left=1.9pt,
  right=1.9pt,
  top=1.9pt,
  bottom=1.9pt]
\textbf{RQ2 Answer:} 
Matched \qiskit-\cirq configurations yield systematically different fingerprints quantified beyond bootstrap uncertainty and consistent with output divergence on \bench{} programs.
\end{tcolorbox}

\section{Conclusion and Outlook}
\label{sec:agenda}
Lightweight empirical noise signatures open new QSE tooling directions, supporting noise-aware debuggers and testing frameworks, aiding compiler optimisation, and educational artefacts, necessary for mature QSE ecosystems. We took the first step toward cross-platform portability analysis by generating a fingerprint for each simulator--backend--noise configuration and quantifying deviations from ideal expectations. Empirically, we observe gaps between configurations across ecosystems, providing developers a concrete diagnostic artefact to reason about when a quantum program may perform well under one configuration yet degrade when ported to another, and alternatively, near-identical fingerprints imply (at the measurement resolution) low portability risk.
Future work will expand to additional platforms and hardware backends, refine estimation precision, integrate fingerprints into development and testing tools, and expand the probe set with additional observables and measurement bases to capture a wider range of noise effects and produce richer fingerprints.  
\looseness=-1

\vspace{0.2 cm}
\noindent\textbf{Acknowledgement. }
Authors are listed in alphabetical order. We thank CloudLab \cite{cloudlab2019} for providing the infrastructure that enabled our evaluation. Note: This paper is an extended preprint version of paper \cite{4a8dd8fae115401383f8834a2f67fe22}.

\bibliographystyle{IEEEtran}
\bibliography{references}
\end{document}